# Measurement of minute volumes of chiral molecules using in-fiber polarimetry

Florian Schorn,[1,2] Arabella Essert, [3] Yu Zhong, [4] Sahib Abdullayev,[4] Kathrin Castiglione, [3] Marco Haumann,[4*] Nicolas Y. Joly [5,2,1*]

[1] Max-Planck-Institut für die Physik des Lichts, Erlangen
[2] Friedrich-Alexander-Universität Erlangen-Nürnberg (FAU), Interdisziplinäres Zentrum für Nano-strukturierte Filme
[3] Friedrich-Alexander-Universität Erlangen-Nürnberg (FAU), Lehrstuhl für Bioverfahrenstechnik (BVT)
[4] Friedrich-Alexander-Universität Erlangen-Nürnberg (FAU), Lehrstuhl für Chemische Reaktionstechnik (CRT)
[5] Friedrich-Alexander-Universität Erlangen-Nürnberg (FAU), Department Physik

*Corresponding authors: nicolas.joly@fau.de; marco.haumann@fau.de

**ABSTRACT:**

We report an opto-fluidic method that enables to efficiently measure the enantiomeric excess of chiral molecules at low concentration. The approach is to monitor the optical activity induced by a Kagome-lattice hollow-core photonic crystal fiber filled with a sub-µl volume of chiral compound. The technique also allows monitoring the enzymatic racemization of R-mandelic acid.

R- and S-enantiomers of a chiral molecule are non-superimposable mirror images from each other. While their chemical and physical bulk properties are mostly identical [1], their biological effect often differs drastically [2]. For this reason, the analysis of chiral molecules plays an important role in e.g. the food, cosmetics and in pharmaceutical industries. [1], [2]. A well-known example here is limonene. R-limonene can be extracted from citrus fruits and has an orange smell. It is used in both food and cosmetics industry. Its counterpart, the S-enantiomer on the other hand smells like turpentine [3]. In the pharmaceutical sector, active components are often applied in small doses of a few µg to a few g per day [4]. As a result, the production capacity is usually much lower than for other chemical compounds. At the same time the purity of pharmaceutical products is subject to strict requirements, since already small doses of impurities can have a harmful effect. This also applies to chiral molecules, since the difference between two enantiomers can completely change the biological impact [2]. An example is the drug Thalidomide, which was prescribed to treat anxiety and insomnia. In this case the harmful substance is the S-enantiomer of the pharmaceutical active R-Thalidomide itself. While the R-enantiomer is a very reliable sleeping drug, the S-enantiomer is highly teratogenic, causing malformations in unborn children. In the case of Thalidomide, these side effects cannot be completely avoided even by using a pure enantiomer, because the other one can nevertheless be formed by an in vivo racemization reaction [2], [5]. In order to avoid such problems for newly developed drugs in the future, it is necessary to carefully monitor their production process.

In conventional chemical processes, both enantiomers are often produced in equal parts, as racemate. However, due to their almost identical chemical and physical properties, it is very difficult to separate the two enantiomers with purely physical methods. The existing separation methods are mainly based on different forms of chiral chromatography [6]. In order not to waste the undesired enantiomer, it is also often recycled using a racemization reaction. A more elegant approach, on the other hand, would be enantio-selective reactions in which the desired enantiomer can be produced in a targeted manner. One way of doing this is to modify the steric demand of the homogeneous catalyst [7], [8]. Another way is the use of biological processes, such as enzymes or bacteria, which often have a very high enantioselectivity [9]. Regardless of the way enantio-pure molecules are produced or separated from each other, the purity must be carefully analyzed. Different enantiomers of chiral molecules are distinguished by measuring their optical activity. This consists of analyzing the rotation of a linearly polarized beam propagating through a sample by an angle $\Delta\alpha$:

$$\Delta\alpha = \alpha(ee_x) - \alpha_{racem} = \alpha_\lambda^T \cdot L \cdot c \cdot M \cdot ee_x \quad (1)$$

where the enantiomeric excess $ee_x$ describes the ratio of the R- to S- enantiomers:

$$ee_R = \frac{c_R - c_S}{c_R + c_S} = -ee_S \quad (2)$$

and $\alpha_{racem}$ corresponds to the angle measured for the racemate case when both enantiomers have equal concentration. L is the optical pathlength, c the total concentration of the chiral component with the molar mass M. According to the convention, only the ee of the excess enantiomer is used such that ee is always positive. The specific rotation of the chiral component $\alpha_\lambda^T$ depends on both the temperature T and the wavelength λ, according to the Biot equation [10]

$$\alpha_\lambda^T = \frac{A(T)}{\lambda^2} \quad (3)$$

where A(T) is a fitting function of the temperature. Optical activity is commonly measured at a wavelength of 589 nm and at ambient temperature. By operating at shorter wavelengths, the specific rotation can be increased by a factor 2 to 5. However, this may also lead to undesired UV-induced interactions. On the other hand, the temperature dependence A(T) highly depends on the molecular structure [11]. To improve the sensitivity of the measurement, increasing the pathlength through the sample does not present any particular physical issue. It is however always accompanied by an increase in the required liquid volume. Analyzing pharmaceutical compounds can then become extremely expensive, the production costs for these are often more than a 100 times higher than for common bulk chemicals [9].

Here we describe the use of hollow-core photonic crystal fiber (HC-PCF) as an optofluidic alternative to the conventional 1 cm wide cuvette. An HC-PFC consists of a periodic arrangement of glass capillaries that surround the empty core region. The entire HC-PCF can serve as a container for the fluid under study while maintaining guidance of the analyzing light in its core region (Fig. 1).

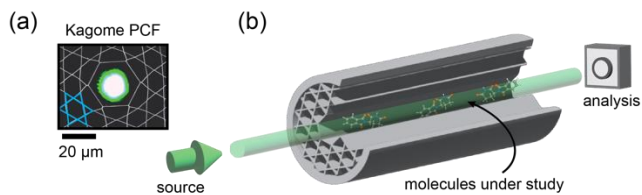

Fig. 1: (a) Micrograph of the Kagomé-lattice HC-PCF used for the experiment. Concept of the experiment. The unit-cell of the Kagomé lattice cladding is indicated in blue. The near field of the output end of the fiber, recorded with a CCD camera, is shown on the SEM of the fiber. (b) Concept of the experiment. The molecules that fill the entire length of the HC-PCF are probed by the laser beam. Analysis include polarization, mode profile, spectrum.

The inset of Fig. 1 shows a micrograph of the Kagomé-lattice hollow-core photonic crystal fiber used for the current work. The silica parts appear in grey while the dark region are the channels that are filled with a fluid. While the entire fiber can be filled with liquid or gas, the light can still be guided in its core region, as long as the refractive index of the fluid is lower than that of the glass. The near-field of the output of the fiber is superimposed on the micrograph of the fiber in Fig. 1. As the result, such broadband waveguides are an ideal platform for any kind of light-matter interaction and it allows many different methods of analysis [12], [13] . In the case of optical activity, the signal strength can easily be improved by a factor of 50 to 100 by simply increasing the fiber length, while keeping the volume of the sample minute. As we reported recently, the fiber can also be used as a platform for in-situ analysis of chemical reactions [14].

The experimental setup is shown in Fig. 2. All experiments were carried out at room temperature. A Kagome-type HC-PCF with a core size of 23 μm was used for the experiments. The characteristic star-of-David unit cell of the Kagome-lattice fiber is indicated in blue on the micrograph of the fiber on Fig. 1. The maximal fiber length that we used is 70 cm. The two ends of the fibers were clamped inside custom-made liquid cells. The dead volume of one of the liquid cells was ca. 50 μl. To fill the fiber, the liquid was first placed in a pressure vessel and then pressurized with nitrogen gas at 12 bar. At first, we flushed the residual air out of the fiber by opening the valves of both liquid cells. In a second step, the bypass valve was closed while keeping one liquid cell opened. The flow through the fiber can be stopped by opening the bypass valve in order to equalize the pressure on both sides. The pump consists of a 532 nm continuous wave laser diode system with an output power of 50 mW. We ensure that any residual ellipticity of the polarization prior the fiber is eliminated by using a λ/4-waveplate (Fig. 2). The orientation of the linearly polarized light is adjusted with a λ/2-waveplate prior the coupling lens. The laser was coupled into the fiber core using an aspheric lens (f = 25 mm). The fiber output was sent to a polarimeter (Thorlabs, PAX1000VIS/M).

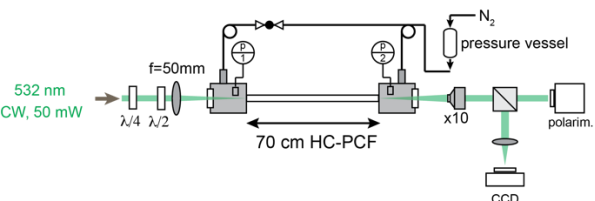

Fig. 2: Experimental setup. A Kagome fiber (SEM on Fig. 1) is clamped with both ends into two liquid cells. The solution to be analyzed is filled into a pressure vessel and pressed into the fiber with 12 bar N2 gas. The pump is a 532 nm CW microchip laser. A set of waveplates control the input polarization at the input of the fiber. The polarization after the fiber is analyzed with a polarimeter. A CCD camera monitored the near field of the fiber.

As a proof of principle experiment, we first measured the specific rotation of commercially available chemicals. Starting from pure 2-butanol (RS: Carl Roth, >98.5%; R: Acros organics, >99%; S: Alfa Aesar, >98%) dissolved in methanol, we filled three lengths of hollow-core PCFs with R-2-butanol with different enantiomeric excess. The value of the measured optical activity over the enantiomeric excess is shown in Fig. 3. All measurement points were averaged over a time span of 10 s. The measured value for pure racemic was used as a reference and subtracted from all measured points. As expected from Eq. (1) we observed a clear linear dependence on the effective interaction length as well as on the concentration of the chiral compound.

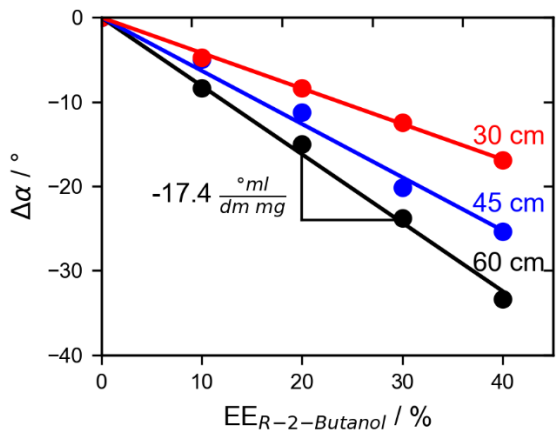

Fig. 3: Absolute measured optical activity for pure 2-butanol over enantiomeric excess for different fiber length, points are measured values, the lines correspond to Eq. (1). T = 20 °C; λ = 532 nm (continuous wave)

Using Eq. (1), we extracted the specific rotation $\alpha_{\lambda=532nm}^{T=20°}$ for R-2-butanol of -17.1 ± 0.2 °ml g$^{-1}$ dm$^{-1}$. By correcting the wavelength with the Biot equation (Eq. (3)), this value is in very good agreement with the manufacturer's value of -17.5 ± 0.9 ° ml g$^{-1}$ dm$^{-1}$ at 532 nm (-14.3 ± 0.7 ° ml g$^{-1}$ dm$^{-1}$ for 589 nm)[15], [16]. In the current case, the hollow-core fiber is 60 cm offering a 60x gain in comparison to conventional cuvette, while requiring a sample volume of less than 1 µl. The total volume contrasts with the internal volume of a conventional cuvette that is between 200 µl and 2 ml. This could be further reduced through further optimization of the liquid cells and the filling mechanism, while the path length could also be further increased through optimization of the guidance properties.

Since chiral chemicals are rarely used as highly concentrated pure substances in pharmaceutical industry, we carried out an additional series of measurements with 2-butanol diluted in methanol. The results for both R- and S-2-butanol are shown in Fig. . Since the absolute value of the specific rotation of the two enantiomers is expected to be identical, but with opposite signs, the absolute value of the optical activity is shown here for better comparability. For each concentration we measured the output polarization state for 5 different initial orientations of the input linear polarization. The mean value of the optical activity of these five measurements is used in the plot and the absolute standard deviation from the mean value is displayed as error bars. The highest standard deviation that we measured is 1.7 °, which is more than 7 times higher than the polarimeter's accuracy of 0.25 °. We attributed this error to the presence of higher-order spatial modes propagating in the hollow-core fiber. Although these modes will all be affected in the same manner by the optical activity of the compound, their respective contribution is likely to change due to in-coupling perturbation and to external stress on the fiber resulting in a random modification of the output polarization. Between each measurement we monitored the output mode and adjust the in-coupling so as to limit the contribution from higher-order modes as much as possible.

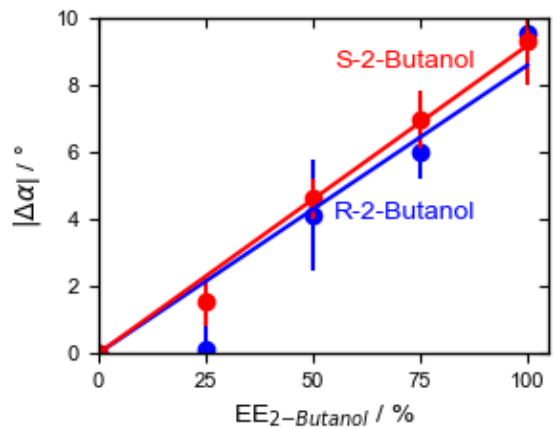

Fig. 4: Measured optical activity for R- and S-2-butanol dissolved in methanol over enantiomeric excess; points are measured values, the lines correspond to Eq. (1); $c_{2-But, tot}$ = 0.6 M, L = 70 cm; T = 20 °C; λ = 532 nm (continuous wave)

Despite the presence of higher-order modes propagating in the fiber, the measurement is significantly more precise than the conventional method because of the long interaction length that we used. From the data presented in Fig. 4 we calculated a specific rotation $\alpha_{532}^{T}$ for R-2-Butanol of -16.5 ± 3.2 ° ml g$^{-1}$ dm$^{-1}$ and +17.7 ± 2.6 ° ml g$^{-1}$ dm$^{-1}$ for S-2-Butanol. As expected, the less concentrated the compound the smaller the rotation of the linear input polarization and the larger the uncertainty. However, both values, although slightly different, are very close to the value given by the manufacturer.

We validated our method by evaluating $\alpha_{\lambda=532}^{T}$ for R-mandelic acid and R-limonene. The Table 1 present the results of our measurements. In each case we obtained a very good agreement with the known data.

Table 1: Comparison of measured and expected optical activities for different chiral components; limonene and 2-butanol: 1 M in methanol; mandelic acid: 0.6 M in water; *estimated with equation (3) using the value for 589 nm reported by the manufacturer

|  | $\alpha_{532}^{20}$/ ° ml g$^{-1}$ dm$^{-1}$ | |
| --- | --- | --- |
|  | Measured | Literature |
| R-2-butanol | -16.5 ± 3.2 | -17.5 [15], [16]* |
| S-2-butanol | +17.7 ± 2.6 | +17.5 [15], [16]* |
| R-mandelic acid | -173.3 ± 2.1 | -187.5 [17]* |
| R-limonene | +144.3 ± 2.6 | +145.1 [18] |

In a second step we use our technique to monitor the evolution of the enantiomeric excess during the enzymatic racemization of R-mandelic acid (RMA). In this reaction we use mandelate racemase from Pseudomonas putida ATCC

12633 as the catalyzing enzyme. It was produced as described by Golombek and Castiglione [19] with minor modifications: For strep-tag II affinity chromatography, a binding buffer containing 10 mM MOPS, pH 8, 150 mM NaCl, 3.3 mM $MgCl_2$, 0.01 % w/v bovine serum albumin (BSA) was used, supplemented with additional 50 mM D-biotin for elution. The purified enzyme was then desalted by PD-10 columns packed with Sephadex G-25 resin (Cytiva) and stored at -80 °C after freeze-drying in liquid nitrogen. The storing buffer is composed of the binding buffer containing 25 mM instead of 150 mM NaCl. For the reaction experiments 100 ml of a basic solution were mixed. It consists of 1 M MOPS buffer (Carl Roth, >99%) with a pH value of 7.5, 25 mM NaCl (Carl Roth, >99%), 3.3 mM $MgCl_2$ (Carl Roth, >99%), and 0.005 w. % BSA (Carl Roth, >98%). From this basic solution 5 ml was used to create a diluted enzyme stock solution with an enzyme concentration of 6.4 µg ml$^{-1}$. This solution was stored at 4 °C. For each experiment 9 ml of a 0.111 M R-mandelic acid solution was created out of the basic solution as well. The reaction was initiated by adding 1 ml of the enzyme stock solution to the mandelic acid solution, resulting in a final enzyme concentration of 0.64 µg ml$^{-1}$ and a final mandelic acid concentration of 0.1 M. Since the concentration of all other components remained equal in both mixtures the procedure ensured that the initial concentrations were always the same. We neglected the small differences in mixing volumes.

We realized the racemization reaction at room temperature. The results for the evolution of ee during the reaction are shown in Fig. 5. The stated age corresponds to the time that the enzyme was stored as diluted stock solution at 4 °C before its use. As expected, the presence of enzymes yields the decay of the enantiomeric excess. It can be expected that at the end of the racemization reaction, both concentrations of the R- and S- enantiomers are equal and the optical activity due to the chiral chemicals should fall to zero. The two solutions are mixed and then immediately filled into the fiber. At the end of the procedure, the fundamental spatial mode of the fiber was adjusted by imaging the near field at the output of the fiber. The filling of the fiber and the optimization of the propagating mode can take up to 10 minutes during which the reaction is already taking place. The impact of this initial deadtime used for filling the fiber and optimizing the propagating mode can be estimated using the Michaelis-Menten-equation for the initial reaction rate with the turnover number $k_{cat}$ = 1124 s$^{-1}$ and the Michaelis constant $K_m$ = 1.1 mM [20]. In our case, a value of 90 to 80% for ee$_{RMA}$ is expected at the beginning of the measurement. While this is the case for one of our experiments (red curve on Fig. 5), we see that the other experiment starts at a much lower enantiomeric excess of around 60%. Since the ee is directly measured through the evaluation of the rotation angle, contribution from higher-order modes may be the reason of the discrepancy with the expected value of ee (t = 10 mn). Although we optimize the in-coupling by visualizing the near-field at the output of the fiber, the modal content from one experiment to the next may vary. By contrast with the previous experiments, where ee is fully known and fixed allowing repetition of the same experiment so as to reduce experimental errors, we cannot modify the experimental conditions during the entire duration of the reaction. Another significant difference is that we worked with ~10x less concentration of mandelic acid in order not to damage the enzymes because of the pH value. As a result of this the sensitivity of the measurement is significantly reduced in comparison to our previous experiments. By using the standard deviation of 1.7 ° that we evaluated in our previous experiments, we estimated 6.4 % uncertainty for the enantiomeric excess due to the lower concentration. This is compatible with the observed fluctuations (Fig. 5).

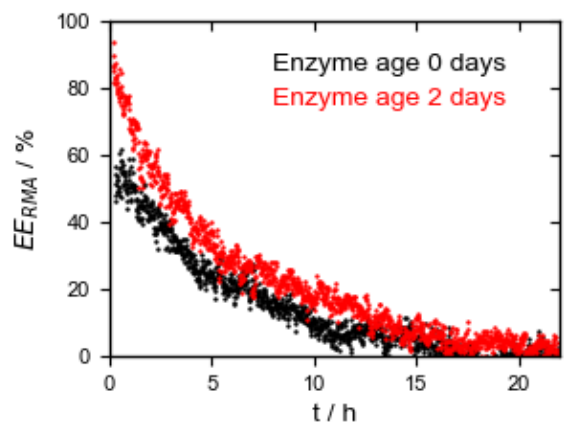

Fig. 5: Measured enantiomeric excess over reaction time for a R-mandelic acid / mandelate racemase solution, Enzyme concentration: 0.64 µg ml$^{-1}$; mandelate concentration: 0.1 M; L = 100 cm, T = 20°C, λ = 532 nm (continuous wave)

In conclusion we showed that we can retrieve the optical activity of chiral molecules by measuring the rotation of the input linear polarization of a beam propagating in a Kagome-lattice hollow-core filled with the enantiomer under study. The required volume of chiral chemical is only 0.5 µl, which is orders of magnitude lower than for conventional methods. Although we use minute volumes, the optical activity of the chiral molecule is significantly improved due to the long interaction length along the optical fiber. Additionally, our procedure allows live-monitoring of the dynamical change of the ee during a racemization reaction. Such a tool could be useful to analyze and optimize the processes involved in racemization reaction. We do not doubt that this technique could also be used during reaction aiming at purifying one specific enantiomer. This could serve efficiently for the development of new drugs in pharmaceutical industry. However, we need to improve the fiber itself in order to reduce the influence of different spatial modes and to further reduce the noise of the measurement. Single-mode hollow-core photonic crystal fibers such as single-ring fiber [21] could significantly improve the quality of the measurement since they are capable of removing all high-order modes from the fiber.

AUTHOR INFORMATION

Corresponding Authors

Nicolas Joly:
Friedrich-Alexander-University Erlangen-Nürnberg, Interdisciplinary Center for Nanostructured Films and Department of Physics; Max-Planck-Institute for the Science of Light, Erlangen, Germany
email: nicolas.joly@fau.de

Marco Haumann: Friedrich-Alexander-Universität Erlangen-Nürnberg (FAU), Erlangen, Lehrstuhl für Chemische Reaktionstechnik (CRT), Germany;
email: marco.haumann@fau.de

Authors

Florian Schorn:
Max-Planck-Institute for the Science of Light, and Friedrich-Alexander-University Erlangen-Nürmberg, Interdisciplinary Center for Nanostructured Films, Erlangen, Germany

Arabella Essert
Friedrich-Alexander-Universität Erlangen-Nürnberg (FAU), Erlangen, Lehrstuhl für Bioverfahrenstechnik (BVT), Germany;

Yu Zhong
Friedrich-Alexander-Universität Erlangen-Nürnberg (FAU), Erlangen, Lehrstuhl für Chemische Reaktionstechnik (CRT), Germany

Sahib Abdullayev
Friedrich-Alexander-Universität Erlangen-Nürnberg (FAU), Erlangen, Lehrstuhl für Chemische Reaktionstechnik (CRT), Germany

Kathrin Castiglione
Friedrich-Alexander-Universität Erlangen-Nürnberg (FAU), Erlangen, Lehrstuhl für Bioverfahrenstechnik (BVT), Germany;

**Abreviations**

| | |
|---|---|
| HC | hollwo core |
| RMA | R-mandelic acid |
| PCF | photonic crystal fiber |

**Acknowledgments**

FS gratefully acknowledges the financial support from the International Max Planck Research School for the Physics of Light in Erlangen.

MH gratefully acknowledges financial support from the Deutsche Forschungsgemeinschaft (DFG, German Research Foundation) – Project-ID 431791331 – SFB 1452 (CLINT).